\documentclass[aps,twocolumn,amsmath,amssymb,nofootinbib]{revtex4}
\usepackage{graphics} 
\usepackage{authordate1-4}
\begin{document}

\title{Cooperativity in Two-State Protein Folding Kinetics} 
\author{Thomas R.\  Weikl$^{1,2,} $\footnote{email: Thomas.Weikl@mpikg-golm.mpg.de}, Matteo Palassini$^{1,} $\footnote{Present address: Laboratoire de Physique Th\'eorique et Mod\`eles Statistiques, B\^at 100, Universite Paris-Sud, 91405 Orsay - France; email: matteo@ipno.in2p3.fr},   and Ken A.\ Dill$^{1,}$\footnote{email: dill@maxwell.compbio.ucsf.edu} \\
{\it\small $^1$Department of Pharmaceutical Chemistry, 
University of California, San Francisco,\\[0cm] 
\small California 94143-2240, USA\\
\small $^2$Max-Planck-Institut f\"ur Kolloid- und Grenzfl\"achenforschung, 14424 
Potsdam, Germany }} 
\noaffiliation

\begin{abstract}
\vspace{0.5cm}

We present a solvable model that predicts the folding kinetics of two-state proteins from 
their native structures. The model is based on conditional chain entropies. It assumes 
that folding processes are dominated by small-loop closure events that can be inferred from 
native structures.  For CI2, the src SH3 domain, TNfn3, and protein L, the model reproduces 
two-state kinetics, and it predicts well the average $\Phi$-values for secondary structures. The barrier to folding is the formation of predominantly local structures such as helices and hairpins, which are needed to bring nonlocal pairs of amino acids into contact.  

\vspace{0.5cm}

\noindent
{\bf Keywords\\}
Protein folding kinetics; two-state folding; folding cooperativity; $\Phi$-value analysis; effective contact order; loop-closure entropy; master equation 

\end{abstract}

\maketitle

\section{Introduction}

Protein folding kinetics is usually modeled in either of three ways. First, there are 
mass-action models that capture the amplitudes and decay rates of the exponentials in the 
folding or unfolding relaxation process \cite{ikai71,tsong71,dill97,englander00}. 
Mass-action models are useful for cataloging the different types of kinetic behavior, but give no insight into molecular structures or mechanisms. Such models do not predict other 
experimental properties, such as $\Phi$-values. Second, there are all-atom or lattice model 
simulations that can explore sequence-structure relationships (see, e.g., 
\cite{duan98,shea01,daggett02}). They are usually limited by computational power to short 
time scales and to studying restricted conformational ensembles. Third, between these 
macroscopic and microscopic extremes, another type of model has recently emerged. This 
class of models uses knowledge of the native structure to infer the sequences of folding 
events \cite{munoz99,alm99,galzitskaya99,debe99, 
shoemaker99,clementi00,hoang00,li01,portman01,ivankov01,alm02,klimov02,flammini02,bruscolini02,bruscolini03,micheelsen03}. Some of these models define partially folded states with one or two contiguous sequences of native-like ordered residues \cite{munoz99,alm99,alm02,galzitskaya99}. Others are based on a Go-model energy function that enforces the global stability of the native state \cite{clementi00,hoang00,li01}. 

We describe here a folding model of the third type. Our model uses knowledge of the native structure to predict the kinetics. However, it differs from previous models in several respects. First, our model focuses on chain entropies and estimates loop lengths from the graph-theoretical concept of effective contact order ECO (see below). We follow time sequences of loop-closure events because we expect that these events reveal how the kinetics is encoded in the native structure. We assume that folding proceeds mostly through closures of small loops, and that large-loop closures are much slower and less important processes. Second, our model focuses on {\it contacts} within the chain, not on whether {\it residues} are native-like or not \cite{munoz99,alm99,galzitskaya99}, because we think the formation of contacts is a more physical description of the folding process.
Therefore, in our model partially folded states are characterized by formed contacts, not by contiguous stretches of native-like ordered residues as in other simple models.
Third, the folding kinetics is described by a master equation that can be solved directly for the macrostates considered here, without stochastic simulations such as molecular dynamics or Monte Carlo.  Hence the present treatment can handle the full spectrum of temporal events.

The present work is related to a recent model of protein zipping 
\cite{firstPaper,secondPaper}. Our fundamental units of protein structure are {\it contact 
clusters}. A contact cluster is a collection of contacts that is localized on a contact map, 
corresponding roughly to the main structural elements of the native structure. Examples of 
contact clusters are turns, $\alpha$-helices, $\beta$-strand pairings, and tertiary pairings 
of helices. A central quantity in our models is the effective contact order 
(ECO)\cite{fiebig93,dill93}. The ECO is the length of the loop that has to be closed in order 
to form a contact, given a set of previously formed contacts or contact clusters. The 
premise is that the formation of the {\em nonlocal} contact clusters requires the prior 
formation of other, more {\em local}, clusters. 

Our model predicts average $\Phi$-values 
for secondary structural elements that are in good agreement with the experimentally 
observed values for several two-state proteins.
 It shows that $\Phi$-value distributions can be understood from 
loop-closure events that are defined by the native topology of a protein. The importance of 
topology for routes and $\Phi$-values has also been previously noted by other groups 
\cite{munoz99,alm99,alm02,clementi00,vendruscolo01}. 

To compute the dynamics, we use a master equation. Several previous studies of the folding
kinetics of lattice heteropolymer models have also been based on master equation methods 
\cite{leopold92,chan93,cieplak98,ozkan01,ozkan02,ozkan,schonbrun}. These methods have the advantage that they require no {\it ad hoc} assumptions about what the transition 
state is. The transition state emerges in a direct physical way from the solution to the 
master equation. However, the lattice models are too simplified to treat specific amino acid sequences or specific protein structures.  Lattice models focus on transitions between {\it microstates}, the individual chain conformations, since these are the fundamental units of structure in such models. Our present master equation describes transitions between {\it macrostates}, defined by the contact clusters of a given protein structure. In this way, the present model aims to make  closer contact with experiments. 

\section{The Model}

\subsection*{Contact clusters}

To compute the folding kinetics, we start with the native contact map, the matrix in which 
element $(i,j)$ equals 1 if the residues $i$ and $j$ are in contact, and equals 0 otherwise. 
Two residues are defined as being in contact if the distance between their $C_{\alpha}$ or 
$C_{\beta}$ atoms is less than 6 Angstroms. 
%

\begin{figure}
\vspace*{-4.5cm}

\hspace*{-3.8cm} 
\resizebox{2\linewidth}{!}{\includegraphics{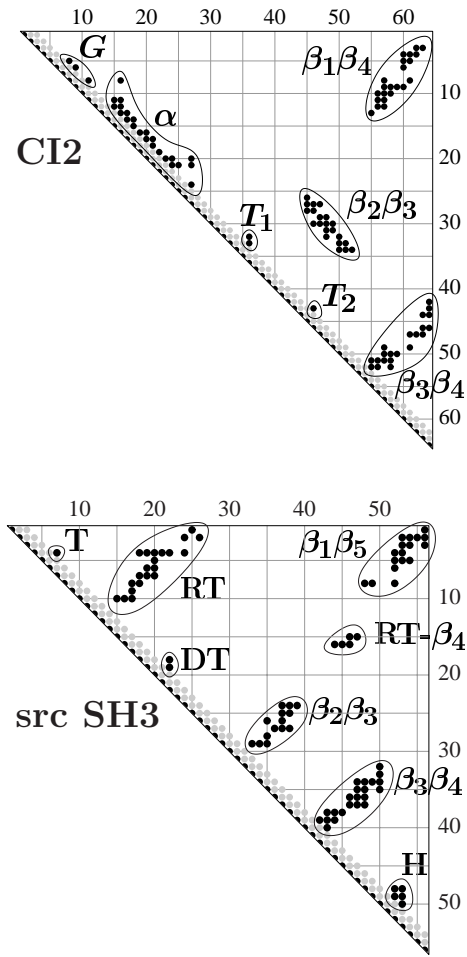}} 
\vspace{-7.2cm} 
\caption{Native contact maps and contact clusters for CI2 and the src SH3 domain. 
\label{clusters}} 
\end{figure}

Next, we divide the native contact map into contact clusters. Each contact cluster 
corresponds to a structural element of the protein. Two contacts $(i,j)$ and $(k,l)$ are 
defined as being in the same cluster if they are close together on the contact map, 
according to the distance criterion that $|i-k|+|j- l|\le 4$. We define two types of clusters: 
local and nonlocal. Clusters are {\it local} if they contain at least one local contact $(i,j)$ having 
contact order $CO = |i- j|\le 6$. Local clusters include helices, turns, or $\beta$-hairpins, 
for example. A cluster is {\it nonlocal} if it has no local contacts; examples include 
$\beta$-strand pairings other than hairpins, and the tertiary interactions of helices. To 
qualify as nonlocal, a cluster must also have more than two contacts; isolated nonlocal 
contacts are not considered to be clusters. Similarly, we do not consider as contributing to 
clusters any `peripheral' contacts $(i,j)$ with a minimum distance 
$|i-k|+|j-l|= 4$ to the other contacts in the cluster.
In general, typical contact maps have only a few isolated nonlocal or 
peripheral contacts. Fig.~\ref{clusters} shows examples of clusters, specifically for 
chymotrypsin inhibitor 2 (CI2) and the src SH3 domain. By our criteria, CI2 has 5 local 
clusters and 2 nonlocal clusters ($\beta_2\beta_3$ and $\beta_1\beta_4$), and the src 
SH3 domain has 6 local clusters and 2 nonlocal clusters (RT-$\beta_4$ and 
$\beta_1\beta_5$). 

\subsection*{States and free energies}

We assume that each cluster is either formed or not; we neglect partial degrees of 
formation. Thus, for a protein with $M$ clusters, there are $2^M$ possible states that 
describe the progression to the native state. Each of these macrostates is characterized by a 
vector $n=\{n_1,n_2,\ldots,n_M\}$, where $n_i = 1$ indicates that cluster $i$ is formed 
and $n_i = 0$ indicates that cluster $i$ is not formed. 

In our model, the free energy of the protein as a function of the state $n$ of cluster 
formation is given by: 
\begin{equation}
F_n = \sum_{i=1}^M n_i\left[c\cdot \ell_i(n)+ f_i\right] \label{stateFs} 
\end{equation}
Each cluster $i$ that is formed ($n_i=1$) contributes to the free energy $F_n$ of the state 
$n$ with two terms: A state-dependent free energy of loop closure $c \cdot \ell_i(n)$ 
(`initiation' free energy), and a free energy $f_i$ for forming the cluster contacts 
(`propagation' free energy). Here, $c$ is a loop-closure parameter. The quantity 
$\ell_i=\ell_i(n)$ is the {\it initiation ECO} \cite{firstPaper} for cluster $i$. The initiation 
ECO of a cluster is the length of the smallest loop that must be closed in order to form that cluster from the other existing clusters. For a local cluster, the initiation ECO is the smallest CO among the contacts. For a nonlocal cluster, the initiation ECO depends on 
the presence of other clusters in the state $n$. 

In general, the initiation ECO also depends on the sequence through which those clusters are formed. However, in order to apply the master equation formalism, we need a free energy and thus we require a definition of initiation ECO that is only a function of state.  For this purpose, we use the following scheme. If only one nonlocal cluster is formed in a certain state, the initiation ECO of that cluster is the smallest ECO among the cluster contacts, given all the local clusters formed in that state. If multiple nonlocal clusters are present in a state, we consider all the possible sequences along which these clusters can form, and determine the one having the smallest sum of ECOs. For instance, for a state with two nonlocal clusters 
$C_i$ and $C_j$, there are two sequences: (1) $C_i$ $\to$ $C_j$, and (2) $C_j$ $\to$ 
$C_i$. The minimum ECOs for the clusters are determined sequentially: $\ell_i^{(1)}$ and 
$\ell_j^{(1)}$ along sequence (1), and $\ell_i^{(2)}$ and $\ell_j^{(2)}$ along sequence 
(2). If $\ell_i^{(1)} + \ell_j^{(1)}$ is smaller than $\ell_i^{(2)} + \ell_j^{(2)}$, the 
initiation ECOs $\ell_i$ and $\ell_j$ of the clusters $i$ and $j$ in the given state are taken 
to be $\ell_i^{(1)}$ and $\ell_j^{(1)}$. The initiation ECOs $\ell_i$ and $\ell_j$ are an 
estimate for the smallest loop lengths required to form the two clusters in the state. 

In eq.~(\ref{stateFs}), the free energy cost of the loops is estimated by a simple linear approximation in the loop length. This is not unreasonable since the range of relevant ECOs only spans roughly one order of magnitude, from about $\ell \simeq 3$ to $\ell \simeq 30$ or 40. In general, determining the free energy of a chain molecule with multiple constraints or contacts is a complicated and unsolved problem. For the simpler problem of hairpin-like loop closures, several estimates have been given in the literature (see, e.g., \cite{chan90,galzitskaya99,ivankov01}).

In principle, this model could treat the detailed energetics of each folding route, if each of 
the $M$ clusters were characterized by its own free energy $f_i$. But here we consider a 
simpler version of the model. We assume that there are only two parameters for the free 
energy of formation: $f_i=f_{l}$ for propagating any local cluster, and $f_i=f_{nl}$ for 
propagating any nonlocal cluster. To obtain two-state folding and agreement with 
experimental $\Phi$-values, we find that $f_l$ must be nonnegative and $f_{nl}$ must be 
negative. This is consistent with the experimental observation that local structures, such as 
helices or $\beta$-hairpins, are generally unstable in isolation. Similar in spirit, the diffusion-collision model of Karplus and Weaver assumes that microdomains, e.g.~helices, are individually unstable \cite{karplus76,karplus94}. Thus, the rate-limiting barrier to folding in our model turns out to be the formation of mostly local structures needed to reduce 
the ECOs of nonlocal clusters. The driving force for overcoming this barrier is the 
favorable free energy $f_{nl}$ of assembling the nonlocal clusters.
%

\begin{figure}
\resizebox{\linewidth}{!}{\includegraphics{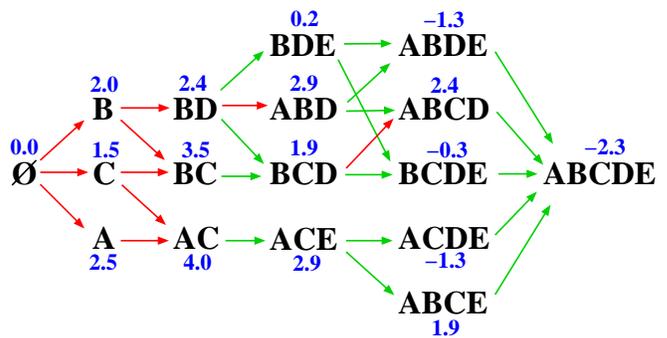}}
 \caption{Energy 
landscape for the src SH3 domain as a function of the 5 major clusters (A) RT, (B) 
$\beta_2\beta_3$, (C) $\beta_3\beta_4$, (D) RT- $\beta_4$, and (E) $\beta_1\beta_5$. 
Here, BD, for example, means that only clusters B and D are formed. The free energies 
given by eq.~(\ref{stateFs}) are shown in blue (the units are $k_B T$). Red arrows indicate 
uphill steps in folding direction, green arrows downhill steps. For clarity, states with free 
energies larger than 4 $k_B T$ are neglected. \label{landscape}} 
\end{figure}

The predicted free energy landscape of the src SH3 domain is shown in 
Fig.~\ref{landscape}, using the parameters $f_l=0$ and $c=0.5$ $k_B T$, where $k_B T$ is 
Boltzmann's constant $\times$ temperature. The value of $f_{nl}$ is chosen so that the 
equilibrium probability that the two nonlocal clusters RT-$\beta_4$ and $\beta_1\beta_5$ 
are both folded (`native state') is 0.9,  which gives $f_{nl}=-6.6$ $k_B T$ for src SH3. 
With these parameter settings, we obtain a good agreement with average experimental $\Phi$-values for the src SH3 domain and other two-state folders (see below).
For clarity, we show in the figure only a reduced set of states based on the 5 major clusters $RT$, $\beta_2\beta_3$, $\beta_3\beta_4$, RT-$\beta_4$, and $\beta_1\beta_5$. The three small clusters T, DT, 
and H have negligible effects on the folding kinetics and on the $\Phi$-values. Only states
 differing by the formation of a single cluster are kinetically connected.
The uphill steps in this model either are steps in which a local cluster is formed, or steps 
involving high ECOs. The downhill steps are steps in which a nonlocal cluster is formed with a low ECO, or steps in which a local cluster significantly reduces the ECOs of previously formed nonlocal clusters. The model predicts two main folding routes. Along the upper route (E) $\beta_1\beta_5$ folds after (D) RT-$\beta_4$; along the lower route, they form in the opposite order. Along these routes, the barriers (highest free energies states) are the states in which two clusters are formed: BD and BC for the upper route, and AC for the lower route. 

\subsection*{Master equation}

In this section, we describe the folding dynamics. We use the master equation, 
\begin{equation}
\frac{\text{d} P_n(t)}{\text{d} t} = \sum_{m\neq n} \left[ w_{nm} P_{m}(t) - 
w_{mn}P_n(t)\right] , 
\end{equation}
which gives the time evolution of the probability $P_n(t)$ that the protein is in state $n$ at 
time $t$. Here, $w_{nm}$ is the transition rate from state $m$ to $n$. The master 
equation can be written in matrix form 
\begin{equation}
\frac{\text{d}{\boldsymbol P}(t)}{\text{d} t} = -{\boldsymbol W} \boldsymbol{P}(t) 
\end{equation}
where $\boldsymbol{P}(t)$ is the vector with elements $P_n(t)$, and the matrix elements 
of $\boldsymbol{W}$ are given by 
\begin{equation}
W_{nm} = -w_{nm} \hspace{0.3cm}\text{for}\hspace{0.3cm} n\neq m; \hspace{0.5cm} 
W_{nn} = \sum_{m\neq n} w_{mn} . 
\end{equation}
The transition rates are given in terms of the free energies by 
\begin{equation}
w_{nm}=\frac{\delta_{|n-m|,1}}{t_o} \left[1+\exp\left(\frac{F_n-F_{m}} {k_B 
T}\right)\right]^{-1} \label{transrates} 
\end{equation}
where $t_o$ is a reference time scale. The only transitions that are assigned to have 
nonzero rates $w_{nm}$ are those incremental steps that change the state $n$ by a single 
cluster unit. This is enforced by the term $\delta_{|n- m|,1}$ in eq.~(\ref{transrates}) 
where the Kronecker $\delta_{i,j}$ is  one for $i=j$ and zero otherwise. 
The condition $|n-m|=1$ is only satisfied by pairs of states $n=\{n_1,\ldots,n_M\}$ and 
$m=\{m_1,\ldots,m_M\}$ with $n_k\neq m_k$ for a single cluster $k$, and with 
$n_k=m_k$ for all other clusters. The transition rates (\ref{transrates}) satisfy detailed balance, $w_{nm} P_{m}^e = w_{mn} P_{n}^e$ where $P_{n}^e\sim \exp[-F_n/(k_B T)]$ is the equilibrium weight for the state $n$. We have chosen here the `Glauber dynamics' with $w_{nm} \sim (1+\exp[(F_n-F_m)/(k_B  T)])^{-1}$. Another standard choice satisfying detailed balance is the Metropolis dynamics, which should lead to equivalent results.  

The detailed balance property of the transition rates implies that the eigenvalues of the matrix $\boldsymbol{W}$ are real. One of the 
eigenvalues is zero, corresponding to the equilibrium distribution, while all other 
eigenvalues are positive \cite{vanKampen}. The solution to the master equation is given by 
\begin{equation}
\boldsymbol{P}(t)=\sum_{\lambda} c_{\lambda} \boldsymbol{Y}_{\lambda} \exp[-\lambda 
t] \label{pdis} 
\end{equation}
where $\boldsymbol{Y}_\lambda$ is the eigenvector corresponding to the eigenvalue 
$\lambda$, and the coefficients $c_\lambda$ are determined by the initial condition 
$\boldsymbol{P}(t=0)$. For $t\to\infty$, the probability distribution $\boldsymbol{P}(t)$ 
tends towards the equilibrium distribution $\boldsymbol{P}^e \sim \boldsymbol{Y}_0$ 
where $\boldsymbol{Y}_0$ is the eigenvector with eigenvalue $\lambda=0$. 

Solving the master equation gives a set of $2^M$ eigenvalues, each with its associated 
eigenvector. Each eigenvalue represents a relaxation rate. As initial conditions at $t=0$, 
we start from the state in which no clusters are formed. This corresponds 
to folding from high temperatures or high denaturant concentrations. 

\section{Results}

\subsection*{The cooperativity in two-state kinetics}

\begin{figure}
\vspace*{-6.7cm}\hspace*{-3cm}
\resizebox{1.7\linewidth}{!}{\includegraphics{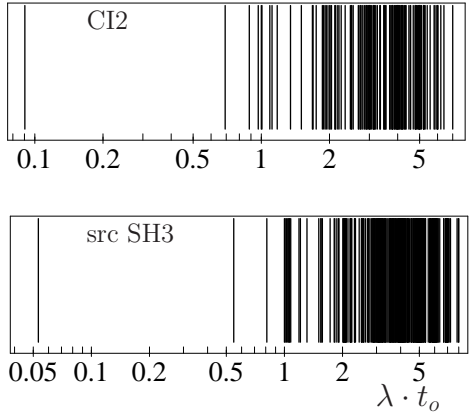}} 

\vspace*{-8cm} 
\caption{Eigenvalue spectra for CI2 and the src SH3 domain in units of 
$1/t_o$ where $t_o$ is the reference time scale for the transition rates (\ref{transrates}). 
\label{spectra}} 
\end{figure}

\begin{figure}
\vspace*{-5cm}\hspace*{-3cm}
\resizebox{1.8\linewidth}{!}{\includegraphics{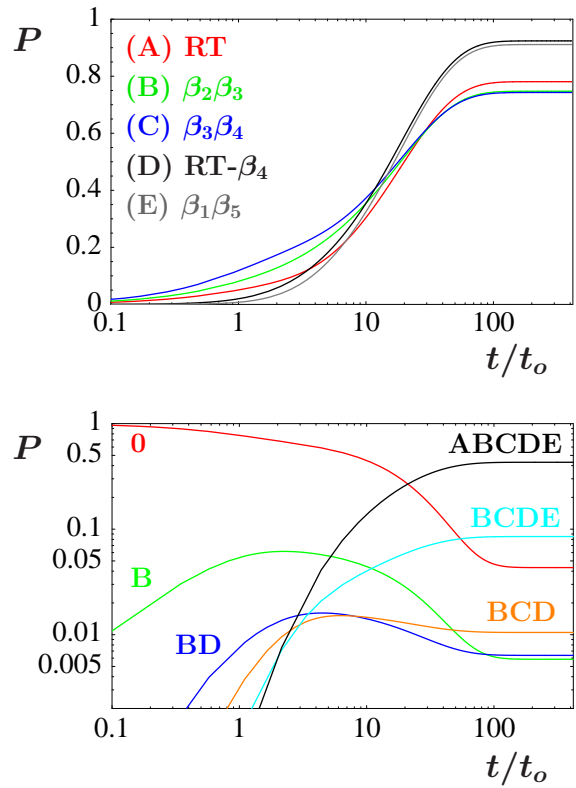}} 

\vspace*{-6cm}

\caption{(Top) Time evolution of the formation probability $P$ for the major {\it clusters} 
of the src SH3 domain during folding (see Fig.~\ref{clusters}). (Bottom) Time evolution of 
{\it state} probabilities for the exemplary path 0 $\to$ B $\to$ BC $\to$ BCD $\to$ BCDE 
$\to$ ABCDE of the src SH3 domain (see also Fig.~\ref{landscape}). The initial state at 
time $t=0$ is the denatured state in which none of the clusters is formed. \label{evolution}} 
\end{figure}

The signature of two-state kinetics is the existence of one slow relaxation process (described by a single exponential), separated in time from $2^M-1$ fast relaxations (a `burst' phase).
Fig.~\ref{spectra} shows  the eigenvalue spectra for CI2 and the src SH3 domain, based on using the parameters $c=0.5$ $k_B T$, local cluster free energy $f_l=0$, and a nonlocal cluster free energy chosen so that the equilibrium `native' population with all nonlocal clusters formed has probability 0.9.  The latter condition leads to $f_{nl}=-7.9$ $k_B T$ for CI2, and $f_{nl}=- 6.6$ $k_B T$ for src SH3. 
Fig.~\ref{evolution} shows the predicted folding dynamics for the src SH3 domain.
%

The spectra in Fig.~\ref{spectra} show that for these proteins, the eigenvalues do indeed separate into a slow single-exponential step and a burst phase, consistent with the experimental observation of two-state behavior. The slowest relaxation rate $\lambda_1$ is about  one order of magnitude smaller than the other nonzero eigenvalues (see Fig.~\ref{spectra}). At times $t \gtrsim 1/\lambda_1$, the probability distribution (\ref{pdis}) is well approximated by $\boldsymbol{P}(t)\simeq c_0\boldsymbol{Y}_0 + c_1 \boldsymbol{Y}_1 \exp[- \lambda_1 t]$ where $\boldsymbol{Y}_0$ is the eigenvector with eigenvalue 0, which characterizes the equilibrium state, and $\boldsymbol{Y}_1$ is the eigenvector with eigenvalue $\lambda_1$. 

The typical time evolution of the folding process predicted by the model is as follows. We have two time scales, $t\simeq t_o$ and $t_F \simeq 1/\lambda_1$. Time $t_o$ is a characteristic of the burst phase in the model and $t_F$ is the single-exponential folding time. At the earliest times, $t < t_o$, single local clusters start to form: examples are the clusters A, B, and C of the src 
SH3 domain, see Fig.~\ref{landscape}. As shown in Fig.~\ref{evolution}, on this time 
scale, each cluster is only weakly populated, with a probability less than  10$\%$.  Any structures having  larger-scale organization -- cluster pairs, triplets, etc.\ -- have negligible populations. At 
intermediate times, $t_F\gtrsim t \gtrsim t_o$, there is a crossover from the burst phase to the 
single-exponential folding process. During these intermediate times, cluster pairs (AC, BC, BD) 
begin to form. Fig.~\ref{landscape} shows that these pairwise clusters are the barrier 
events, i.e., they represent the conformational states of maximum free energy obtained 
during folding. Finally, on the longest time scale, $t\simeq t_F$, the pairwise and triplet 
clusters reach sufficiently high populations to assemble into multi-cluster 
complexes, proceeding downhill in free energy to the native structure.

What is the basis for the cooperativity of folding in our model, i.e.\ for the separation of time scales?  
First, the formation of local structures in our model reduces the loop-closure entropies for the 
formation of the nonlocal structures.  Second, only the nonlocal structures have favorable 
propagation free energies $f_i=f_{nl} < 0$.  Hence, the formation of the nonlocal 
structures stabilizes the overall fold, and thus also the local structures.  
The barrier arises from the positive free 
energies in eq.~(\ref{stateFs}) due to the formation of local structures and loops (see 
Fig.~\ref{landscape}).  Interestingly, if we set the free energies for local structure 
formation to be negative by several $k_B T$, we obtain fast multi-exponential downhill 
folding, without a barrier. Based on experiments and theory, such downhill folding has 
been recently postulated for the protein BBL \cite{garcia02}. 

To understand the cooperative folding in the model, it is instructive to 
turn off the loop-closure term in eq.~(\ref{stateFs}) by setting $c=0$. Then all $M$ clusters 
are independent of each other. In that case, there is no cooperativity. 
It can be shown that the matrix $\boldsymbol{W}$ then has the eigenvalues 
$\lambda=j/t_o$ where $j$ is an integer between 0 and M, the number of clusters.
 Each of these eigenvalues has a 
population that is given by the binomial coefficient $j!/[j!(M-j)!]$. This gives a broad 
non-two-state spectrum. Hence, the separation of time scales -- and the two-state cooperativity --
arise in this model from the coupling of the clusters via the loop-closure term in 
eq.~(\ref{stateFs}). 

To see the magnitude of the barrier, note that the folding rate $\lambda_1$ is related to 
the height of the energy barrier on the folding landscape. For comparison, consider a 
mass-action model with three states D $\leftrightarrow$ T $\leftrightarrow$ N (denatured state, 
`transition' state, native state) and transition rates as in eq.~(\ref{transrates}). The folding rate 
is given, to a very good approximation, by $(1/2) t_o^{-1} \exp[-F_b^\ddagger/(k_BT)]$ for barrier 
energies $F_b^\ddagger=F_T^\ddagger - F_D^\ddagger \gg k_B T$. The factor of 1/2 comes from the fact that a molecule 
in state T can jump both to D and N, with almost equal probability, since both sides of 
high-barrier transition states are steep downhills. Now, for the energy landscape of the src 
SH3 domain shown in Fig.~\ref{landscape}, the minimum barrier has free energy 2.4 $k_B 
T$ for state BD. The corresponding barrier crossing rate of $(1/2) t_o^{-1}\exp[-2.4]$ is in 
good agreement with the folding rate $\lambda_1\simeq 0.05/t_o$ (see 
Fig.~\ref{spectra}). 

Experiments have been interpreted either as indicating that burst phases involve structure 
formation or that burst phases are processes of non-structured polymer collapse, 
depending on the protein and the experimental method 
\cite{englander00,callender98,gruebele98,eaton98,parker00,ferguson03}. In our model, 
the burst phase is a process of structure formation. Non-structured collapse is beyond the scope,
or resolution, of our model, because the model has only a single fully unstructured state -- the state 
in which none  of the clusters is formed. 
 The burst phase in our model captures fast preequilibration events within the denatured state 
in response to initiating the folding conditions at $t=0$. In the model, this denatured state is an ensemble of macrostates on one side of the barrier in the energy landscape (see Fig.~\ref{landscape}).
It is reasonable to assume that such preequilibration events within the denatured state exist also for real proteins.
However, whether these events can be detected as burst phases in experiments should 
depend on the initial conditions, experimental probes, etc.

\begin{table}

Table 1: Maximum probability $P_{max}$ and ${\bf Y}_1$ elements 
for transient states of the src SH3 domain.\\

\begin{center}
\begin{tabular}{c|cc}
\hspace*{0.1cm}state\hspace*{0.1cm}& \hspace*{0.2cm}$P_{max}$\hspace*{0.3cm} & \hspace*{0.2cm}$Y_1$ element \\
\hline 
C & 0.13 & 0.21 \\
B & 0.062 & 0.10 \\
A & 0.047 & 0.08 \\[0.1cm]
\hline 
BD & 0.016 & 0.019 \\
BC & 0.010 & 0.017 \\
AC & 0.007 & 0.011 \\[0.1cm]
\hline 
BCD & 0.015 & 0.010 \\
ABD & 0.004 & 0.003 \\
ACE & 0.004 & 0.001 
\end{tabular}
\end{center}
\end{table}

During folding or unfolding, certain conformations will be populated transiently.  If the 
populations of those conformations are always small, we call them `hidden intermediates' 
\cite{ozkan02}. The population of a hidden intermediate conformation rises to a 
maximum, $P_{max}$, then falls as the protein ultimately becomes fully folded. The term 
`hidden' means that $P_{max}$ is always small enough that it does not contribute an 
additional kinetic phase; i.e., the folding kinetics is two-state. Here, we consider two 
quantities. (1) We compute $P_{max}$ for the transient states. For simplicity, we 
consider only the 5 major clusters $RT$, $\beta_2\beta_3$, $\beta_3\beta_4$, 
RT-$\beta_4$, and $\beta_1\beta_5$. (2) We look at the elements of the eigenvector ${\bf 
Y}_1$, the eigenvector corresponding to the smallest eigenvalue $\lambda_1$. These 
elements show how the various conformations grow and decay with rate $\lambda_1$ as 
folding proceeds. Table 1 shows that the maximum population $P_{max}$ correlates well 
with the elements of ${\bf Y}_1$. For a typical
route of src SH3, Fig.~\ref{evolution} 
(bottom) illustrates the decay of the denatured state and hidden intermediates and the 
growth of the native state, all with rate $\lambda_1$. 

\subsection*{Average $\Phi$-values for secondary structural elements}

The effects of a mutation on the folding kinetics are often explored through experimental 
measurements of a $\Phi$-value, which is defined as
\begin{equation}
\Phi=- \frac{k_B T \ln (k_f^\prime/k_f)}{\Delta G^\prime - \Delta G} \label{phi} 
\end{equation}
where $k_f$ is the folding rate of the native protein and $\Delta G$ is its stability, and 
$k_f^\prime$ and $\Delta G^\prime$ are the corresponding quantities for the mutant 
protein. 

Since the minimal structural units in our model are clusters of contacts, we do not calculate 
$\Phi$-values for single-residue mutations. Rather, we consider whole helices and strands 
as units. To compare with experiments, we average the experimental $\Phi$-values over 
all the residues composing a given secondary structural element. 

To calculate average $\Phi$-values for secondary structures, we consider `mutations' that 
change the free energy $f_i$ of a contact cluster according to 
\begin{equation}
\Delta f_i(j) = x_{ji} \epsilon \label{mut} 
\end{equation}
where $x_{ji}$ is the fraction of residues of the secondary structural element $j$ that are 
involved in contacts of the cluster $i$, and $\epsilon$ is a small energy. For example, if the 
secondary structural element $j$ contains $m_1$ residues, and $m_2\le m_1$ of these 
residues appear in contacts of the cluster $i$, we have $x_{ji}=m_2/m_1$. Note that $0\le 
x_{ji}\le 1$, where the value $x_{ji}=1$ is obtained if the whole secondary structural 
element $j$ has contacts in cluster $i$. Thus the $\Phi$-value for the secondary structural 
element $j$ is given by eq.~(\ref{phi}) with 
\begin{equation}
\ln (k_f^\prime/k_f)=\ln (\lambda_1^\prime/\lambda_1) 
\end{equation}
where $\lambda_1^\prime$ is the smallest nonzero eigenvalue of the mutant with cluster 
free energies $f_i \to f_i + \Delta f_i(j)$, and 
\begin{equation}
\Delta G' - \Delta G= \sum_{i} \Delta f_i(j) 
\end{equation}
For $\epsilon \ll k_B T$, we find that the calculated $\Phi$-values are nearly independent 
of $\epsilon$. We choose here $\epsilon=0.01$ $k_B T$. 

\begin{figure*}
\vspace*{-3cm} \hspace*{-2.6cm} 
\resizebox{0.8\linewidth}{!}{\includegraphics{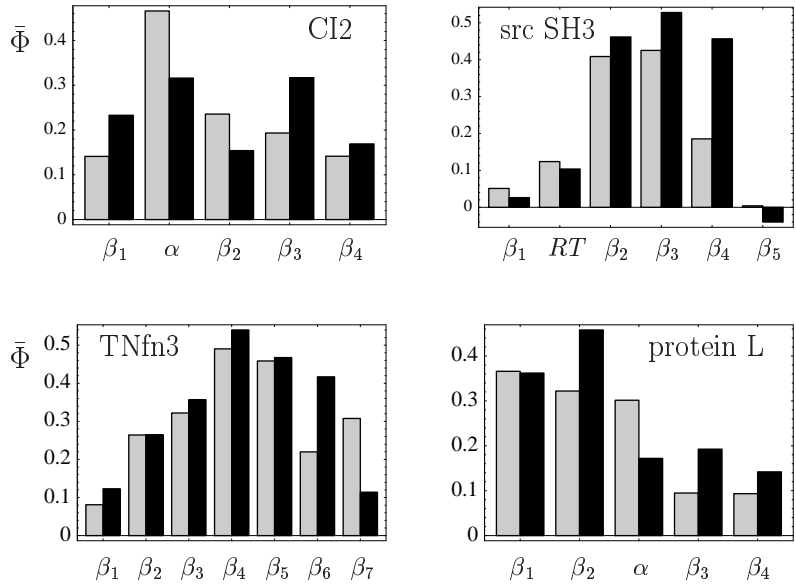}} 

\vspace{-9.3cm} \caption{Theoretical and average experimental $\Phi$-value distributions 
for the secondary structural elements of CI2, the src SH3 domain, TNfn3, and protein L. 
The parameter of the loop closure term is $c=0.5$, and the free energy of the local clusters 
is $f_l=0$. The free energy $f_{nl}$ of the nonlocal clusters is chosen so that the 
probability that all nonlocal clusters are formed is $0.9$ in equilibrium. \label{phisI}} 
\end{figure*}

\begin{figure}[b]
\vspace*{-6cm} \hspace*{-1cm} 
\resizebox{1.8\linewidth}{!}{\includegraphics{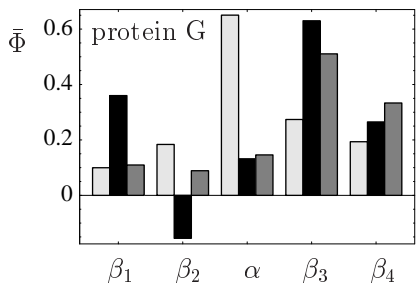}} 

\vspace{-12cm} \caption{Comparison of theoretical and experimental $\Phi$-value 
distributions. (Light grey): theoretical $\Phi$-values for the same parameters as in 
Fig.~\ref{phisI}. (Black): average experimental $\Phi$-values. (Dark grey): theoretical 
$\Phi$-values when assuming that the free energy of the $\alpha$-helix cluster is $f_i=-0.5 
k_B T$, deviating from the standard value $f_l=1.5 k_B T$ for the local clusters. 
\label{phisII}} 
\end{figure}

Predicted $\Phi$-values are compared with experiments in Fig.~\ref{phisI}. The theoretical 
$\Phi$-values were calculated with the same parameters for all four 
proteins (see figure caption).  The predicted values agree well with the experimental 
values. This comparison indicates that the folding kinetics of these proteins is dominated by generic features of the fold topology, rather than by the specific energetic details -- i.e., which residues form contacts, how much hydrogen bonds or hydrophobic interactions are worth, the details of sidechain packing, etc. In the case of protein G (see Fig.~\ref{phisII}), the experimental $\Phi$-value distribution is largely reproduced by making the additional assumption that the $\alpha$-helix cluster has a free energy $f_i=- 2.0$ $k_B T$, rather than the value $f_i=0$ $k_B T$ that we have otherwise used for local clusters (see Fig.~\ref{phisI}). However, even without changing this parameter, the $\Phi$-value distribution reflects the features of the experimental distribution that the $\Phi$-values for the strands $\beta_3$ and $\beta_4$ are larger than those for $\beta_1$ and $\beta_2$. 

\section{Conclusions}

We have developed a simple model of the folding kinetics of two-state proteins. The model 
aims to predict the folding rates of the fast and slow processes, the folding routes, and 
$\Phi$-values for a protein, if the native structure is given. The dominant folding routes are 
found to be those having small ECOs, i.e., steps that involve only small `loop closures'. The 
model parameters include: $c$, an intrinsic free energy for loop closure; $f_{l}$, the free 
energy for propagating contacts in local structures; and $f_{nl}$, the free energy for 
propagating nonlocal contacts. The model predicts that the barrier to two-state folding is the 
formation of local structural elements like helices and hairpins, and that the steps involving their assembly into larger and more native-like structure are downhill in free energy. 

\end{document}